\documentclass[aps,pra,superscriptaddress,reprint,floatfix,longbibliography,showpacs,amsfonts,amsmath,amssymb]{revtex4-1}
\usepackage{graphicx}
\usepackage{hyperref}
\usepackage{xr-hyper}
\usepackage{upgreek}
\usepackage{epstopdf}
\usepackage[note-name=, use-sort-key = false]{notes2bib}
\usepackage{color}
\usepackage{comment}

\def\cm{cm$^{-1}$}

\begin{document}
\title{Influence of magnetic ordering on the optical response of the antiferromagnetic
topological insulator MnBi$_2$Te$_4$}

\author{M. K\"opf}
\affiliation{Experimentalphysik II, Institute of Physics, Augsburg University, 86159 Augsburg, Germany}
\author{J. Ebad-Allah}
\affiliation{Experimentalphysik II, Institute of Physics, Augsburg University, 86159 Augsburg, Germany}
\affiliation{Department of Physics, Tanta University, 31527 Tanta, Egypt}
\author{S. Lee}
\affiliation{Department of Physics, Pennsylvania State University, University Park, PA 16803, USA}
\author{Z. Q. Mao}
\affiliation{Department of Physics, Pennsylvania State University, University Park, PA 16803, USA}
\author{C. A. Kuntscher}
\affiliation{Experimentalphysik II, Institute of Physics, Augsburg University, 86159 Augsburg, Germany}
\email{christine.kuntscher@physik.uni-augsburg.de}

\begin{abstract}
The layered topological insulator MnBi$_2$Te$_4$ has attracted great interest recently due to its intrinsic antiferromagnetic order,
potentially hosting various topological phases. By temperature-dependent infrared spectroscopy over a broad frequency range, we studied the changes in the
optical conductivity of MnBi$_2$Te$_4$ at the magnetic ordering temperature. The temperature dependence of several optical parameters reveals an anomaly at the magnetic phase transition, which suggests the correlation between the bulk electronic band structure and the magnetism. We relate our findings to recent reports on the temperature dependence of the electronic band structure of MnBi$_2$Te$_4$.
\end{abstract}

\maketitle

\section{Introduction}
Ever since the emergence of topology in the field of solid state physics, materials have been classified and characterized with the use of quantum mechanical attributes. As a result, the quantum anomalous Hall insulator has been predicted and discovered as promising application in the field of spintronic or energy-efficient electronics~\cite{Yu.2010,Chang.2013,Chang.2015,Lee.2019}. For the realization of a solid hosting this effect, topological and magnetic properties have to coexist, as magnetic ordering breaks time-reversal symmetry and thus creates dissipationless spin-polarized chiral edge states~\cite{Shi.2019}. Topological insulators (TI) are insulating in the bulk, but possess conducting surface or edge states, which are Dirac fermions protected by inversion and time-reversal symmetry \cite{Hasan.2010,Qi.2011}. Therefore, the impact of magnetic ordering removes the symmetry protection and enables a potential drive into another topological phase, such as quantum anomalous Hall insulator, topological axion insulator or chiral Majorana fermions~\cite{Tokura.2019,Vidal.2019}.

\begin{figure*}[t]
	\includegraphics[width=1\linewidth]{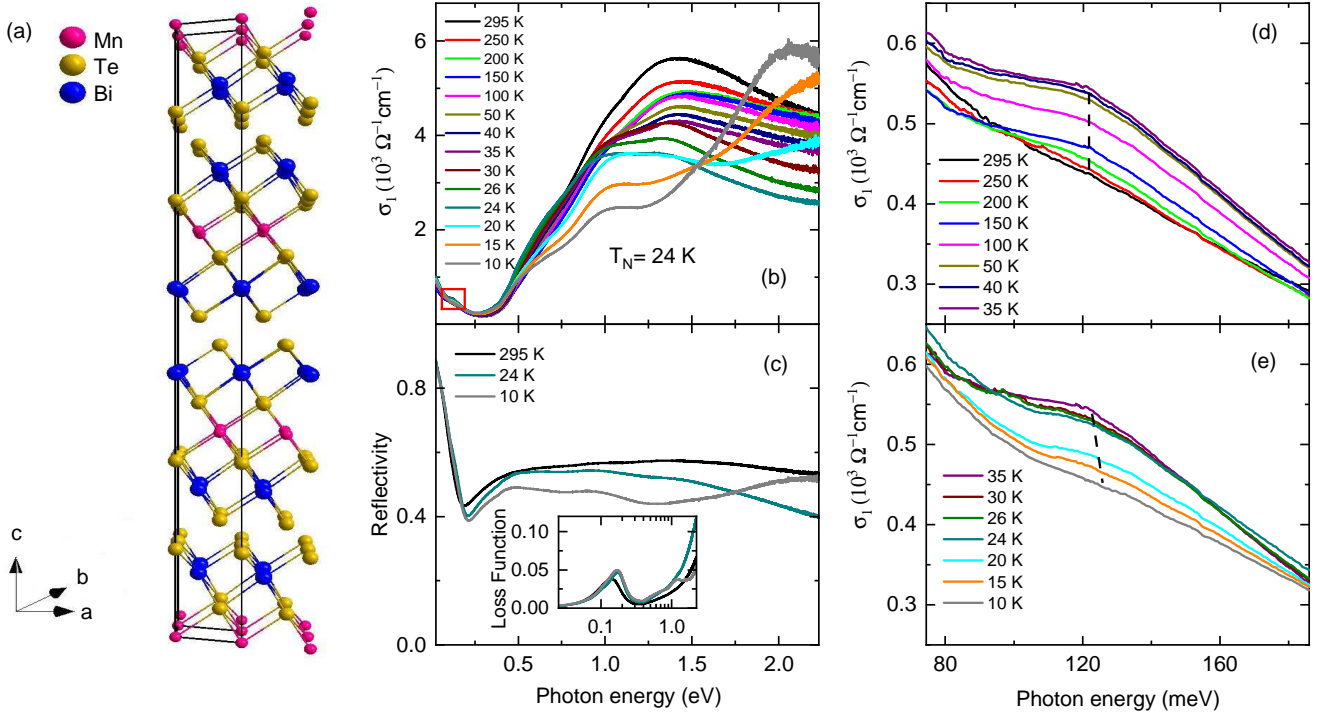}
	\caption{\label{fig.conductivity} {(a) Crystal structure of MBT \cite{Rani.2019}. (b) Optical conductivity $\sigma_1$ of all measured temperatures. The red rectangle marks the zoomed-in frequency range shown in (d) and (e). (c) Reflectivity spectrum for selected temperatures at 295, 24 and 10\,K, together with the corresponding loss function in the inset. (d) and (e) Zoom-in of the optical conductivity $\sigma_1$ in the low-energy range during cooling down from 295 to 35\,K and
from 35 to 10\,K, respectively. Dashed lines highlight the peak position of the excitation.}}
\end{figure*}

MnBi$_2$Te$_4$ (MBT) has been reported as the first intrinsic antiferromagnetic TI \cite{Zhang.2019a,Li.2019b,Otrokov.2019a}.
Both quantum anomalous insulator and axion insulator have been demonstrated in two-dimensional thin layers of MBT \cite{Deng.2020,Liu.2020a}
MBT belongs to the family of ternary chalcogenides with the chemical formula MB$_2$T$_4$ (M = transition metal; B = Bi, Sb; T = Te, Se, S)~\cite{Li.2019b} and crystallizes in a rhombohedral symmetric structure with the space group $R\bar{3}m$.
Fig.\ \ref{fig.conductivity}(a) shows a sketch of the crystal structure with the lattice parameters $a=4.3825$\,\AA\ and $c=42.6849$\,\AA~\cite{Rani.2019}. MBT can be viewed as a naturally formed heterostructure, where a Mn-Te layer is inserted in every Te-Bi-Te-Bi-Te quintuple layer, hence consisting of Te-Bi-Te-Mn-Te-Bi-Te septuple layers with van-der-Waals-type interaction along the $c$ direction.
Due to the insertion of the transition metal Mn$^{2+}$, the MBT is intrinsically magnetic, since the Mn ions with spin $S=5/2$ provide a magnetic moment of $\sim 5\,\upmu_{\mathrm{B}}$ per unit cell~\cite{Deng.2020}. A magnetic ordering sets in below the phase transition at $T_{\mathrm{N}} = 24\,$K~\cite{Hao.2019}, where the compound shows parallel alignment within each of the septuple layers and anti-parallel alignment between the layers. Thus, below $T_{\mathrm{N}}$ MBT is an $A$-type antiferromagnetic (AFM) compound with out-of-plane magnetic moments.

Since the gapping of the topological surface state (TSS) in MBT due to the intrinsic AFM ordering is crucial for the realization of the axion insulating state, the energy dispersion of the TSS has been extensively investigated by angle-resolved photoemission spectroscopy (ARPES) experiments by various groups \cite{Lee.2019,Chen.2019a,Hao.2019}. Contradictory results were obtained regarding the gapping of the TSS: Some studies found a gapping of the TSS for temperatures already far above $T_{\mathrm{N}}$, whereas others revealed the absence of gapping even below $T_{\mathrm{N}}$, i.e., that the TSS is unaffected by the magnetic ordering.
In contrast to the TSS, some hints for the interplay between the bulk electronic band structure and the magnetism in MBT were found by  magnetotransport and ARPES studies \cite{Lee.2019,Chen.2019a}. For example, when cooling below the magnetic phase transition, a splitting of the conduction band was observed, which was attributed to surface ferromagnetism of local magnetic moments of Mn atoms buried in the surface layer \cite{Chen.2019a}.

Here, we apply infrared spectroscopy, which is a more volume-sensitive technique compared to ARPES, to probe the changes in the {\it bulk} electronic band structure of MBT at the magnetic ordering transition. Our results clearly reveal strong changes in the optical conductivity at $T_{\mathrm{N}}$, thus confirming the impact of magnetic ordering on the {\it bulk} electronic properties of MBT.

\section{Methods}
The investigated air-cleaved MBT single crystal with lateral dimensions 1\,mm $\times$ 1.5\,mm was synthesized using the flux growth method~\cite{Lee.2019}. Raman measurements confirmed the positions of the vibrational modes E$^2_g$ and A$^2_{1g}$ (not shown), as found by Li et al.~\cite{Li.2020}. The sample was mounted on a cold-finger micro-cryostat from CryoVac Konti for cooling. We performed reflectivity measurements between 295 and 10\,K in a frequency range from 0.01 to 2.23\,eV  (100 to 18000\,\cm) with a Bruker Hyperion infrared microscope coupled to Bruker Vertex80v FTIR spectrometer. A silver layer was evaporated onto half of the sample surface, which served as reference for obtaining the absolute reflectivity.

The measured reflectivity spectra were extrapolated in the low- and high-energy range by taking into account the dc conductivity~\cite{Li.2020, Cui.2019, Zeugner.2019} and the volumetric data for the x-ray optic file~\cite{Rani.2019}, respectively. Thereby, the Kramers-Kronig relations were applied to transform the reflectivity into optical functions, like the complex optical conductivity $\hat{\sigma}$, the complex dielectric function $\hat{\epsilon}$, and the loss function defined as -Im(1/${\hat{\epsilon}}$). The RefFIT program \cite{Kuzmenko.2005} was used to describe the optical data with the use of a Drude-Lorentz fitting and to observe the relative changes in the spectra with decreasing temperature.


\begin{figure}[t]
	\includegraphics[width=0.9\linewidth]{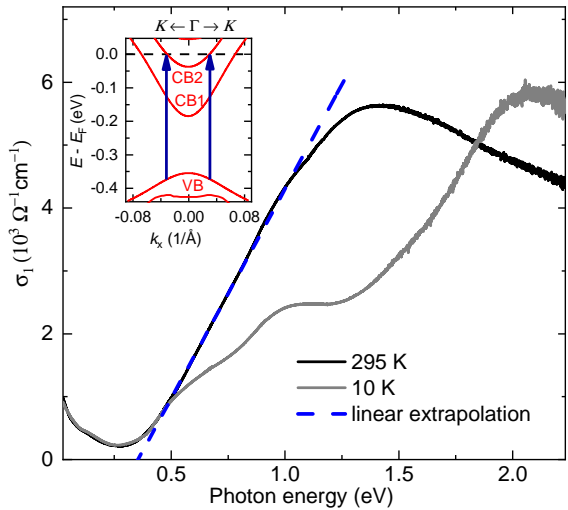}
	\caption{\label{fig.absorption-edge} {Optical conductivity $\sigma_1$ of MBT at 295 and 10\,K, with the onset of the interband transitions between the valence and conduction bands at $\sim$0.3~eV.
The temperature-dependent onset of the interband transition is determined from the zero crossing of the linear fit, as illustrated for the 295 K spectrum.
Inset: Scheme of the {\it bulk} electronic band structure of MBT along the $\Gamma$-$K$ direction of the Brillouin zone according to Chen et al.~\cite{Chen.2019a}, where CB1 and CB2 mark the conduction bands and VB the valence bands. The onset of interband transitions between the VB and CB in the $\sigma_1$ is due to transitions marked by the blue arrows.}}
\end{figure}

\section{Results and Discussion}
The reflectivity spectrum of MBT at 295~K shows a plasma edge at low frequencies, and the plasma minimum is followed by a plateau-like behavior [see Fig.\,\ref{fig.conductivity}(c)]. The corresponding loss function reveals a well-defined plasmon peak, as shown in the inset of Fig.\,\ref{fig.conductivity}(c). The temperature-induced effects in the low-frequency reflectivity are only slight, however, major changes occur in the higher-energy range, especially below $T_{\mathrm{N}}$.


The optical conductivity spectra $\sigma_1$, which were obtained from the reflectivity spectra by Kramers-Kronig analysis, are depicted in Fig.\,\ref{fig.conductivity}\,(b) as a function of temperature. The $\sigma_1$ spectrum at room temperature (see also Fig.\ \ref{fig.absorption-edge}) contains Drude contributions at low frequencies due to intraband transitions, a strong onset of interband transitions above $\approx$0.3~eV, and a broad absorption band centered at around 1.4~eV (see discussion below).
The most pronounced changes in the optical conductivity with temperature occur in the high-energy range, where the $\sigma_1$ first decreases with increasing temperature. Below $T_{\mathrm{N}}$=24\,K, the profile of the high-energy optical conductivity changes drastically, as the level further decreases below 1.5~eV, but rises above. At 10~K, the high-energy optical conductivity contains two hump-like features at $\sim$0.6~eV and $\sim$1~eV and an absorption band at $\sim$2.1~eV. These drastic changes in the $\sigma_1$ spectrum across 24\,K  give an initial hint for the magnetic phase transition, which apparently affects the electronic band structure close to the Fermi energy.

The temperature-induced changes in the low-energy ($<$0.25~eV) range of the optical conductivity are more subtle, as illustrated in Figs.\,\ref{fig.conductivity}\,(d) and (e), which show a zoomed-in region for the spectra for temperatures 295\,K - 35\,K and 35\,K - 10\,K, respectively. Here, an increase of $\sigma_1$ is observed with decreasing temperature from 295 to 35\,K [Fig.\,\ref{fig.conductivity}\,(d)], followed by a decrease of $\sigma_1$ for further cooling from 35 to 10\,K [Fig.\,\ref{fig.conductivity}\,(e)].
Also, one finds a weak excitation in this low-frequency range, whose energy position is marked by the dashed lines in Figs.\,\ref{fig.conductivity}\,(d) and (e): The peak position of this weak excitation is roughly constant during cooling down to $T_{\mathrm{N}}$, and abruptly shifts to higher energies below $T_{\mathrm{N}}$. The origin of this feature will be discussed later.

\begin{figure*}[t]
	\includegraphics[width=1\linewidth]{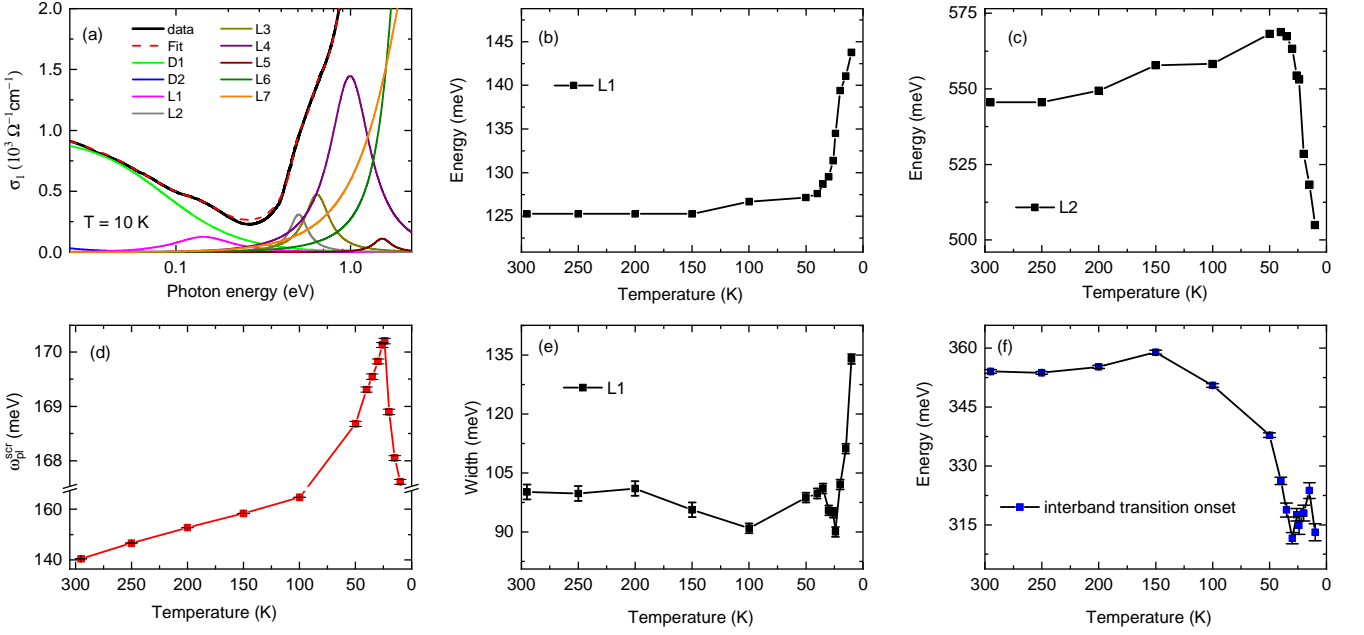}
	\caption{\label{fig.parameters} {(a) Drude D and Lorentz L contributions from the fitting of the optical conductivity of MBT at 10~K. 
Temperature dependence of the (b) position of the L1 oscillator, (c) position of the L2 oscillator, (d) screened plasma frequency $\omega^{\mathrm{scr}}_{\mathrm{pl}}$, as determined from the plasmon peak position in the loss function, (e) width (damping) of the L1 oscillator, and (f) interband transition onset. Please note the interrupted $y$-scale in (d).}}
\end{figure*}

For further investigation of the spectral properties, the onset of the interband transitions
is extracted from the optical conductivity spectrum and compared to the bulk electronic band structure of Ref.\ \cite{Chen.2019a}.
As illustrated in Fig.\ \ref{fig.absorption-edge} for 295~K, the onset of the interband transitions was obtained from the zero crossing of the linear fit of the $\sigma_1$ spectrum, and amounts to 353\,meV (2850\,\cm) at 295~K.
In the inset of Fig.\ \ref{fig.absorption-edge}, the schematic electronic band structure of MBT from Ref.\ \cite{Chen.2019a} is shown:
CB1 and CB2 mark the conduction bands and VB the valence bands, where the electronic states in the vicinity of the Fermi energy have mainly Bi/Te $p$ character (whereas the Mn $d$-bands, which are responsible for the magnetism in MnBi$_2$Te$_4$, are far away
from the band gap) \cite{Li.2019b}.
The onset of interband transitions between the VB and CB in the $\sigma_1$ is due to transitions indicated by the blue arrows, and lies in the energy range 0.3 - 0.4\,eV \cite{Chen.2019a}, in good quantitative agreement with our optical results.
Following the above-described linear fitting procedure, we determined the onset of the interband transitions as a function of temperature, which is depicted in Fig.\ \ref{fig.parameters} (f). Accordingly, the interband transition onset shows a rather weak temperature dependence down to $\sim$50~K, and the determination of its value becomes unprecise close to $T_{\mathrm{N}}$, since the profile of the conductivity close to the transition onset is drastically changing (see the $\sigma_1$ spectrum at 10~K in Fig.\ \ref{fig.absorption-edge} as an example).

For a detailed analysis of the temperature evolution of the optical response, the optical conductivity spectra were fitted with a Drude-Lorentz model containing two Drude terms and six Lorentz terms, as illustrated in Fig.\,\ref{fig.parameters} (a) for 10~K. Hereby, the Lorentz oscillator L7 represents the sum of all high-energetic contributions.
Electric transport studies \cite{Yan.2019a,Lee.2020} showed that the charge carrier type in MBT is mainly electron, with transport mobility $\mu$$\approx$58~cm$^2$/Vs at 2~K  \cite{Lee.2020}. Therefore, we attribute the main, rather broad Drude term D1 to excitations of n-type charge carriers. The scattering rate $\gamma$$\approx$87~meV (700~cm$^{-1}$) of D1 at 10~K corresponds to a scattering time $\tau$$\approx$7.6$\cdot$10$^{-15}$~s, according to $\gamma=1/(2 \pi c \tau)$. From the scattering time, we can calculate the carrier mobility $\mu$ according to $\mu$=$e \tau/m^*$, where $m^*$ is the effective carrier mass which we set equal to the free electron mass,
and obtain $\mu$$\approx$13.4~cm$^2$/Vs, which is a factor of $\sim$3 lower but of the same order of magnitude as the transport mobility \cite{Lee.2020}.

\begin{figure}[b]
	\includegraphics[width=1\linewidth]{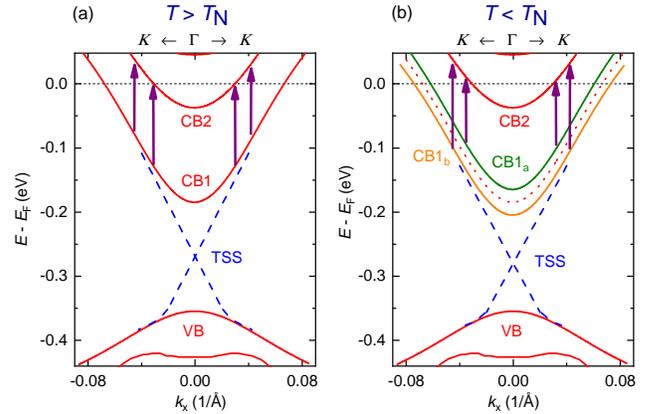}
	\caption{\label{fig.bandstructure} {Scheme of the electronic band structure of MBT along the $\Gamma$-$K$ direction of the Brillouin zone according to Chen et al.~\cite{Chen.2019a}, (a) above and (b) below $T_{\mathrm{N}}$. CB1, CB2, CB1$_{\mathrm{a}}$, and CB1$_{\mathrm{b}}$ mark the conduction bands, while VB indicates the valence band, and TSS the topological surface state.
The red dotted line in (b) indicates the CB1 band above $T_{\mathrm{N}}$, i.e., before its splitting.
Purple arrows indicate the transitions from CB1 to CB2, which is described by the L1 oscillator in Fig.\,\ref{fig.parameters}\,(a,b,e). CB1 splits in two sub-bands (CB1$_{\mathrm{a}}$ and CB1$_{\mathrm{b}}$) at $T_{\mathrm{N}}$, resulting in two energetically close transitions \cite{Chen.2019a}.}}
\end{figure}

By applying this fitting model (two Drude terms, six Lorentz terms) for all temperatures, the position, strength, and scattering rate (width) of each term was obtained.
First, we will discuss the temperature dependence of the position of Lorentzian L1 and its interpretation in terms of the electronic band structure.
According to the schematic electronic band structure above $T_{\mathrm{N}}$, depicted in Fig.\ \ref{fig.bandstructure}(a) \cite{Chen.2019a}, the excitation L1 can be attributed to transitions from the conduction band CB1 to the conduction band CB2, since its energy position is in good agreement with the CB1--CB2 energy difference \cite{Chen.2019a}.
The temperature dependence of the position of the L1 peak is rather weak, but below $T_{\mathrm{N}}$ it abruptly shifts to higher frequencies by $\approx$20~meV (160~cm$^{-1}$) [see Fig.\,\ref{fig.parameters} (b)].
Simultaneously, the width of L1 increases sharply below $T_{\mathrm{N}}$ [see Fig.\,\ref{fig.parameters} (e)], which indicates a significant broadening of this peak.

Also the Lorentzian L2 shows a marked temperature dependence, namely it first shifts to higher frequencies with decreasing temperature, but below $T_{\mathrm{N}}$ it abruptly shifts to lower frequencies by about 74~meV (600~cm$^{-1}$) [see Fig.\,\ref{fig.parameters} (c)].
The L2 peak is located right at the onset of the interband transitions between the valence and conduction bands, and thus the energy position of the peak could serve as an alternative measure of the transition onset, which is affected at temperatures close to $T_{\mathrm{N}}$ (see also Fig.\ \ref{fig.absorption-edge}).

An anomaly at the magnetic phase transition also occurs in the temperature dependence of the
screened plasma frequency $\omega^{\mathrm{scr}}_{\mathrm{pl}}$, which characterizes the position of the plasma edge in the reflectivity spectrum and corresponds to the frequency of the longitudinal plasmon mode.
The value of $\omega^{\mathrm{scr}}_{\mathrm{pl}}$ was obtained from the position of the plasmon peak in the corresponding loss function [see inset Fig.\,\ref{fig.conductivity} (c)] by fitting with a Lorentz function. The so-obtained temperature dependence of $\omega^{\mathrm{scr}}_{\mathrm{pl}}$ is plotted in Fig.\,\ref{fig.parameters} (d): With decreasing temperature down to $T_{\mathrm{N}}$$\approx$24\,K, $\omega^{\mathrm{scr}}_{\mathrm{pl}}$ monotonically increases, but abruptly decreases below $T_{\mathrm{N}}$, forming a sharp maximum.

In summary, the optical response of MBT is strongly affected by the magnetic ordering at $T_{\mathrm{N}}$, as revealed by an anomaly in the temperature dependence of several optical parameters.

Recent angle-resolved photoemission studies on MBT \cite{Chen.2019a} have already suggested that the bulk electronic band structure is affected by the magnetic order setting in below $T_{\mathrm{N}}$.
As illustrated in Fig.\ \ref{fig.bandstructure}, it was found that the conduction band CB1 splits below $T_{\mathrm{N}}$ into two components CB1$_{\mathrm{a}}$ and CB1$_{\mathrm{b}}$, where the first one shifts to higher and the latter one shifts to lower energies by approximately the same energy \cite{Chen.2019a}. Indeed, we find indications for such a splitting in our optical conductivity data:
As already mentioned above, the peak L1 corresponds to transitions between the conduction bands CB1 and CB2.
The width of L1 increases suddenly below 24\,K, representing a peak broadening [see Fig. \ref{fig.parameters}(e)]. This broadening can be attributed to the splitting of the conduction band CB1 into the two contributions CB1$_{\mathrm{a}}$ and CB1$_{\mathrm{b}}$, where the two transitions CB1$_{\mathrm{b}}$$\rightarrow$CB2 and CB1$_{\mathrm{a}}$$\rightarrow$CB2
are very close in energy and thus cannot be resolved in the spectrum. The abrupt shift of L1 to higher energies is another signature of the band splitting, since it corresponds to transitions CB1$_{\mathrm{b}}$$\rightarrow$CB2 occurring below $T_{\mathrm{N}}$, where the CB1$_{\mathrm{b}}$ subband is shifted down in energy with respect to the unsplit CB1.
Overall, our optical results are in good quantitative agreement with the findings from the photoemission experiments \cite{Chen.2019a}.

The correlation between the conduction band splitting and the magnetic ordering was attributed \cite{Chen.2019a} to the exchange interaction between the electronic states and the local magnetic moments of the Mn atoms buried in the surface layer, causing surface ferromagnetism. Since ARPES is a highly surface-sensitive technique, this might seem to be a possible interpretation. In contrast, infrared spectroscopy is more bulk-sensitive
\bibnote{For the dc conductivity $\sigma_{dc}$$\sim$$1\cdot 10^5$~$\Omega^{-1}$m$^{-1}$ of MBT \cite{Li.2020, Cui.2019, Zeugner.2019}
we estimate the skin depth $\delta$$\sim$0.3~$\mu$m for electromagnetic radiation with wavenumber 1000~cm$^{-1}$ according to
$\delta$ = $\sqrt{2/(\sigma_{dc}\cdot \omega \cdot \mu_0)}$ \cite{Fox.2001},
where $\omega$ is the angular frequency and $\mu_0$ the magnetic permeability.}, which on the one hand excludes the observation of the TSS (in contrast to ARPES), but on the other hand ensures its insensitivity to surface magnetism. Therefore, according to our infrared spectroscopy study, the splitting of the conduction band below $T_{\mathrm{N}}$ because of surface ferromagnetism, as proposed by Chen et al.\ \cite{Chen.2019a}, is unlikely. Apparently, a different mechanism must be at play.

\section{Conclusion}
In conclusion, the optical conductivity of MBT, obtained by reflectivity measurements, and its temperature dependence are in good quantitative agreement with recent reports on the bulk electronic band structure.
The optical response of MBT is strongly affected by the magnetic ordering at $T_{\mathrm{N}}$, as revealed by changes in the profile of the optical conductivity and an anomaly in the temperature dependence of several optical parameters. Our findings demonstrate the interplay between the bulk electronic states and the magnetism in MBT.

\begin{acknowledgments}
The financial support for sample preparation was provided by the National Science Foundation through the Penn State 2D Crystal Consortium-Materials Innovation Platform (2DCC-MIP) under NSF cooperative agreement DMR-1539916.
\end{acknowledgments}

\end{document}